\documentclass[10pt,twocolumn,a4paper]{article}
\usepackage{amsmath,amssymb,latexsym}
\usepackage[affil-it]{authblk}

\usepackage{helvet}

\usepackage[textwidth=19cm,textheight=25cm]{geometry}

\usepackage{fancyhdr}
\usepackage{titlesec}
\titleformat*{\section}{\fontsize{12}{20}\bfseries\selectfont}
\titleformat*{\subsection}{\fontsize{10}{17}\bfseries\selectfont}

\usepackage{lastpage}
\usepackage{ifthen}
\fancypagestyle{plain}{ %
  \fancyhf{} 
  \ifthenelse{\value{page}=1}{}{} 
  
}

\usepackage{floatrow}
\usepackage{multirow}

\usepackage{graphicx}
\usepackage{mathptmx}
\usepackage{setspace}

\providecommand{\keywords}[1]{\textbf{\textit{Keywords}} #1}
\usepackage{etoolbox}
\apptocmd{\thebibliography}{\fontsize{9}{0}\selectfont}{}{}%
\begin{document}

\title{\bf A New Approach to Compute the Dipole Moments of a Dirac Electron}
\fancyhf[HLE,HRO]{A New Approach to Compute the Dipole Moments of a Dirac Electron \bf (\thepage\ of \pageref{LastPage})}

\author{Semra Gurtas Dogan%
}
  \author{Ganim Gecim%
}

\author{Yusuf Sucu%
  \thanks{\texttt{ysucu@akdeniz.edu.tr}; Corresponding author}\hspace*{0.1cm}}

\fancyhf[HRE,HlO]{S.Gurtas Dogan, G. Gecim and Y. Sucu  }

\affil{Department of Physics, Faculty of Sciences,\\ Akdeniz University, Antalya, Turkey}
\date{Dated: \today}

\maketitle

\begin{abstract}
We present an approach to compute the electric and magnetic
dipole moments of an electron by using polarization and magnetization parts of
the Dirac current. We show that these dipole moment expressions obtained by our
approach in this study are in agreement with the current experimental results in the literature. Also, we observe that a magnetic field plays an important role in the magnitude of the electrical dipole moment of the electron.
\\
\\
\keywords{Dipole moment, magnetization, polarization, Dirac electron}
\end{abstract}

\section{Introduction}
\label{sec:1}
The electric dipole moment (EDM) and magnetic dipole moment (MDM) are very important
concepts to understand the internal structure of a material and, at the same
time, the intrinsic structure of the elementary particles. Therefore, to
describe the mathematical and physical properties of these concepts, several
studies have been carried out since the beginning of 20th century and they are all summarized in \cite{1}.In
particular, the magnetic dipole moment was considered by various methods
\cite{2,3,4}. In addition, the electric dipole moment has been also investigated
by several other experimental \cite{5,6,7,8,9,10,11,12,13} and theoretical \cite{14,15}
studies. For instance, although predicted value by standard model for the
electric dipole moment is of about $10^{-38}e.cm$, physical models beyond the
standard model report a larger value \cite{16} and references therein.

Dirac showed that an electron in an external electric and a magnetic field has
both an electric and a magnetic dipole moments \cite{17}. The magnetic dipole
moment expressed in his theory of spinning relativistic electron model is
proportional to the spin angular momentum of an electron. The value foreseen
in the model for the magnetic dipole moment is consistent with the
experimental values \cite{18,19}. Also, the electric dipole moment in the same
theory \cite{17} is seen to be proportional to the particle's spin, as in the
magnetic dipole moment.

Dirac theory gives a probabilistic current density for a Dirac particle which
has an electric and a magnetic dipole moment. The Gordon decomposition as a well
known technique used to separate the probabilistic current density into the
densities of convection, polarization and magnetization was performed to
discuss the particle creation in 2+1 dimensional gravity \cite{20}. On the
other hand, in the classical electromagnetic theory, the polarization and
magnetization densities are defined as a magnetic dipole moment and electric
dipole moment per unit volume of an electrical and a magnetic material,
respectively. From this point of view ,these dipole moments can be obtained by integrating the polarization
and magnetization densities  of the Dirac probability current over a hyper surface of the material in the
quantum electrodynamical context, as in classical electromagnetic theory.

The aim of this study is to compute the electric and magnetic dipole moments
of a Dirac electron. For this, we derive the magnetization and polarization
densities using the exact solutions of the Dirac equation in presence of a
constant magnetic field. Thus, the study defines us a new alternative
prescription in obtaining the electrical and magnetic dipole moments of a
relativistic Dirac particle interacting with any flat or a curved background.
With this alternative prescription, we find the magnetic dipole moment of a
Dirac electron to be equal to the Bohr magneton and the electron electric
dipole moment's magnitude is of about $10^{-28} e.cm.$ These values are
consistent with the current experimental results.

The outline of the paper is as follows. Section II includes the exact solutions
of the Dirac equation in a constant magnetic field. In section III, these
solutions are used to write the components of Dirac currents, the
magnetization and the polarization densities, obtained via the Gordon
decomposition. From these densities, we compute the electron magnetic and
electric dipole moment. In section IV, we evaluate and summarize the results of the study.

\section{\label{b1}Dirac particle in a constant magnetic field}

Dirac particle in constant or homogeneous magnetic field is,
naturally, an 2+1 dimensional problem. To discuss the
physical properties of an electron in a constant magnetic field, we
write the covariant form of the Dirac equation in 2+1 dimensional
spacetime:
\begin{equation}
\left\{  i\overline{\sigma}\left[  \partial_{\mu}+ieA_{\mu}\right]
\right\}  \Psi(x)=m_{e}\Psi(x), \label{Equation1}
\end{equation}
where \ $\hbar=c=1$ and $\overline{\sigma}^{\mu}$ is Dirac matrices in 2+1
dimensional, $\overline{\sigma}^{\mu}=(\overline{\sigma}^{0},\overline{\sigma
}^{1},\overline{\sigma}^{2})$ and $\overline{\sigma}^{0}=\sigma^{3}%
,\ \ \overline{\sigma}^{1}=i\sigma^{1},\ \ \overline{\sigma}^{2}=i\sigma^{2},$
which $\sigma^{1}$, $\sigma^{2}$ and $\sigma^{3}$ are Pauli matrices,
$\partial_{\mu}$ is derivative according to $x=\left(  x^{0},\ x^{1},\ x^{2}\right)$ coordinates, and $e$ is charge of the Dirac
particle. Also, $A_{\mu}$ is 3-vector of the electromagnetic
potential in 2+1 dimensions. $\Psi(x)$ is a Dirac spinor with two components,
$\chi$ and $\varphi,$ which are called positive and negative energy
eigenstates, respectively, and $m_{e}$ is the mass of Dirac particle \cite{20}.
To discuss Dirac particle in an external constant magnetic field, we choose $A_{\mu},$ 3-vector
of electromagnetic potential, as follows;

\begin{eqnarray}
A_{0}&=&0,\ \ A_{1}=A_{x}=-\frac{B_{0}}{2}y,\nonumber\\
A_{2}&=&A_{y}=\frac{B_{0}}{2}x, \label{Equation3}
\end{eqnarray}
and define the general wave function of an Dirac electron or Dirac spinor
in terms of positive and negative energy eigenstates as $\Psi(x)=e^{-iEt}\binom{\chi}{\varphi}$ because the problem of the Dirac electron in
a constant magnetic field is stationary in time. Using the explicit
form of the Pauli matrices, Eq.(\ref{Equation1}) is reduced to two
coupled first order differential equations system;
\begin{align}
\left(  E-m_{e}\right)  \chi-\left[  \dfrac{\partial}{\partial x}
-i\dfrac{\partial}{\partial y}-ie\frac{B_{0}}{2}y+e\frac{B_{0}}{2}x\right]
\varphi &  =0,\nonumber\\
\left(  E+m_{e}\right)  \varphi+\left[  \dfrac{\partial}{\partial x}
+i\dfrac{\partial}{\partial y}-ie\frac{B_{0}}{2}y-e\frac{B_{0}}{2}x\right]
\chi &  =0.\quad\label{Equation4}%
\end{align}
After that, as writing Eq.(\ref{Equation4}) in polar coordinates $(r,\theta)$ and
letting $\chi(\rho,\theta)=e^{ik\theta}\rho^{\frac{1}{2}}F_{1}\left(\rho\right)$ and $\varphi
(\rho,\theta)=e^{i\left(  k+1\right)\theta}\rho^{\frac{1}{2}}F_{2}\left(\rho\right)$ by means of the separation of
variable method and by defining a new independent variable, $\rho,$ as
$\rho=\frac{eB_{0}}{2}r^{2}$, we get Eq.(\ref{Equation4}) as follows;
\begin{align}
\left[  \dfrac{d}{d\rho}-\dfrac{k}{2\rho}-\frac{1}{2}\right]  F_{1}
(\rho)+\frac{\left(  E+m_{e}\right)  }{\sqrt{2eB_{0}\rho}}F_{2}(\rho)  &
=0,\nonumber\\
\left[  \dfrac{d}{d\rho}+\dfrac{k+1}{2\rho}+\frac{1}{2}\right]  F_{2}
(\rho)-\frac{\left(  E-m_{e}\right)  }{\sqrt{2eB_{0}\rho}}F_{1}(\rho)& =0.
\label{Equation6}
\end{align}
And, the solutions of these equations are obtained as
\begin{equation}
\Psi=%
\begin{matrix}
e^{-(iEt-ik \theta+\rho/2)}\rho^{\frac{k+1}{2}}\binom{N_{1}\text{ }_{1}
F_{1}\left[  -n,\text{ }k+1,\rho\right]  }{N_{2}e^{i\theta}\rho\text{ }
_{1}F_{1}\left[  -n,\text{ }k+2,\rho\right]  }, \label{Equation7}
\end{matrix}
\end{equation}
where $k\ $\ is the quantum number of the angular momentum, $n$ is the
principal quantum number, $N_{1}$ and $N_{2}$ are normalization constants.
From  Eq.(\ref{Equation6}) and Eq.(\ref{Equation7}), we find the following
relation between $N_{1}$ and $N_{2}$,
\begin{equation*}
N_{1}=\frac{(k+1)\sqrt{2eB}}{(E-m_{e})}N_{2},
\end{equation*}
and determine the energy eigenvalues as follows;
\begin{equation}
E^{2}=m_{e}^{2}+2eB_{0}(n+k+1). \label{Equation8}
\end{equation}

The solutions of the Dirac equation in presence of a constant magnetic field
has been already discussed in the literature \cite{21,22}. However, we rederive the solutions of
the Dirac electron in presence of a constant magnetic field to compute the
electron EDMs and electron MDMs.

To calculate the EDMs and MDMs from the polarization and
magnetization densities, we need asymptotic expressions of the Dirac
electron wave function because contributions to the dipole moments
mainly come from the boundaries. So, it is sufficient to normalize
the asymptotic expressions of the wave function to obtain the
normalization constants, $N_{1}$ and $N_{2}.$ Then, the asymptotic
form of the wave function becomes
\begin{equation}
\Psi\rightarrow N_{2}e^{-(iEt-ik \theta+\rho/2)}\binom{\frac{(k+1)\sqrt{2eB}
}{(E-m_{e})}\rho^{\delta-1/2}}{e^{i\theta}\rho^{\delta}} \label{Equation9}
\end{equation}
where $\delta=\frac{E^{2}-m_{e}^{2}-(k+1)eB_{0}}{2eB_{0}}=n+\frac{k+1}{2}
$ \cite{23} and also the normalization
constant, $N_{2},$ are calculated as
\begin{equation*}
N_{2}^{2}=\frac{eB_{0}\text{ }}{2\pi\Gamma(2\delta)}\left(\frac
{2eB_{0}(k+1)^{2}}{(E-m_{e})^{2}}+(2\delta)\right)^{-1}.
\end{equation*}

\section{\label{b2}The dipole moments of an electron}

We start by writing the Dirac currents in 2+1 dimensions. Using
Gordon decomposition, the current of the Dirac particle is separated
into the convective part, two polarization and one magnetization
parts in 2+1 dimensions \cite{20}, as different from 3+1 dimensions \cite{29}.
The decomposed current in 2+1 dimensions is
\begin{align}
J^{\mu}  &  =\frac{1}{2m_{e}}\left(  \overline{\Psi}\overline{\sigma}
^{\mu\upsilon}\left(  t,r\right)  \Psi\right)  _{,\upsilon}\nonumber\\
&  -\frac{1}{2m_{e}}\overline{\Psi}\left(  \frac{i}{2}g^{\mu\upsilon
}\overleftrightarrow{\partial_{\upsilon}}-qA^{\mu}\right)  \Psi\nonumber\\
&  -\frac{i}{4m_{e}}\overline{\Psi}\left[  \overline{\sigma}^{\upsilon}\left(
t,r\right)  ,\overline{\sigma}_{,\upsilon}^{\mu}\left(  t,r\right)  \right]
\Psi\nonumber\\
&  -\frac{i}{2m_{e}}\overline{\Psi}\left[  \overline{\sigma}^{\upsilon}\left(
t,r\right)  \Gamma_{\upsilon},\overline{\sigma}^{\mu}\left(  t,r\right)
\right]  \Psi\nonumber\\
&  -\frac{i}{4m_{e}}\overline{\Psi}\left[  \overline{\sigma}_{,\upsilon
}^{\upsilon}\left(  t,r\right)  ,\overline{\sigma}^{\mu}\left(  t,r\right)
\right]  \Psi, \label{Equation11}%
\end{align}
where $\overline{\Psi}$ is hermitian conjugate of the Dirac spinor
$\Psi$ and equals to
$\overline{\Psi}=\Psi^{\dagger}\overline{\sigma}^{0}=\Psi^{\dagger
}\sigma^{3}$.

The Dirac current can also be separated in terms of the time, $0$,
and spatial, $l,m=1,2$, components, respectively, as follows;
\begin{equation*}
J_{0}=\partial_{l}\mathbf{P}_{l0}+\rho_{convective}
\end{equation*}
and
\begin{equation*}
J_{l}=\partial_{0}\mathbf{P}_{0l}+\partial_{m}\mathbf{M}_{\left[  ml\right]
}+J_{l\text{ }convective},
\end{equation*}
where $\mathbf{P}_{l0}$ are polarization densities and $\mathbf{M}_{\left[
ml\right]}$ is magnetization density and their explicit forms are given as
\begin{equation}
\mathbf{P}^{l0}=\frac{1}{2m_{e}}\overline{\Psi}\overline{\sigma}^{l0}
\Psi\label{Equation12}
\end{equation}
and%
\begin{equation}
\mathbf{M}^{[lm]}=\frac{1}{2m_{e}}\overline{\Psi}\overline{\sigma}^{lm}\Psi,
\label{Equation13}%
\end{equation}
respectively, $\overline{\sigma}^{0l}=i/2\left[\overline{\sigma}^{0},\overline{\sigma}^{l}\right]$ and $\overline{\sigma
}^{lm}=i/2\left[\overline{\sigma}^{l},\overline{\sigma}^{m}\right]$ [22].
Using these relations, we can calculate the total polarizations and
magnetization, respectively, on the hyper surface, $d\Sigma_{kl}$=$d\Sigma_{0}=d^{2}x$, as
\begin{equation}
p^{l}=\int\mathbf{P}^{0l}d\Sigma_{0} \label{Equation14}
\end{equation}
and
\begin{equation}
m=\int\mathbf{M}^{12}d\Sigma_{12}. \label{Equation15}
\end{equation}
To compute the total polarization and magnetization densities, we insert the
expression of wave function, Eq.(\ref{Equation9}), and its complex conjugate
in Eq.(\ref{Equation12}) and Eq.(\ref{Equation13}), respectively, and
obtain the densities as follows;
\begin{align}
\mathbf{P}_{0}^{1}  &  \approx\frac{N_{2}^{2}\text{ }e(k+1)\sqrt{2eB_{0}}
}{2m_{e}(E-m_{e})}\rho^{2\delta-1/2}e^{-\rho}(e^{i\theta}+e^{-i\theta
}),\nonumber\\
\mathbf{P}_{0}^{2}  &  \approx\frac{N_{2}^{2}\text{ }e(k+1)\sqrt{2eB_{0}}
}{2m_{e}(E-m_{e})}\rho^{2\delta-1/2}e^{-\rho}(e^{i\theta}-e^{-i\theta
}),\nonumber\\
\mathbf{M}^{0}  &  \approx\dfrac{N_{2}^{2}\text{ }}{2m_{e}}\left(
\frac{(k+1)^{2}(2eB_{0})}{\left(  E-m_{e}\right)  ^{2}}\rho^{2\delta-1}
+\rho^{2\delta+1}\right)  e^{-\rho}. \label{Equation16}
\end{align}
Using the Eq.(\ref{Equation14}), Eq.(\ref{Equation15}) and
Eq.(\ref{Equation16}), the electric and magnetic dipole moments of the Dirac
electron are computed, respectively, as follows;
\begin{align}
p^{1}  &  =0\nonumber\\
p^{2}  &  =0\nonumber\\
m  &  =\dfrac{e\hbar}{2m_{e}c}, \label{Equation17}
\end{align}
where we insert $\hbar,$ Planck constant, and $c$, velocity of light.
We observe that the computing procedure directly gives a well-known
expression for the magnetic dipole moment of the electron, as Bohr
magneton, but the electrical dipole moment components vanish because of the axial symmetry stemmed from $\theta$ coordinate. However, if we choose the $\theta$ interval, as a topological deficit of the spacetime,
$\theta\in\lbrack0,2\pi)$ instead of $\theta\in\lbrack0,2\pi]$, which leads to conical singularity, then, the
electrical dipole moment components are obtained as
\begin{equation}\label{eq2}
\left.\begin{aligned}
p^{1}& =\frac{e\text{ }\hbar\varepsilon}{2\pi
m_{e}c\sqrt{2eB_{0}\hbar c}}\\
\phantom{\quad}& \phantom{a} \textnormal{x}\frac{(k+1)\Gamma(2n+k+\frac{3}{2})}{\left(  (k+1)^{2}+  \frac{\varepsilon^{2}%
}{{2eB_{0}\hbar c}} (2n+k+1)\right)  \Gamma(2n+k+1)},\\
p^{2}&=ip^{1}
\end{aligned}\right.
\end{equation}\label{Equation18}
where $\varepsilon=(E-m_{e}c^{2})$ and it can be interpreted as the kinetic energy of the electron.
The EDM of the Dirac electron exhibits an interesting behaviour according to
the relation between $\varepsilon$ and $\sqrt{2eB_{0}\hbar c}$. In the case of
$\varepsilon\gg\sqrt{2eB_{0}\hbar c}$, the electron electric dipole moment (eEDM) expression
in Eq.(\ref{Equation18}) is reduced the following form,
\begin{equation*}
p^{1}\simeq e\lambda_{C}\frac{\sqrt{2eB_{0}\hbar c}}{\varepsilon},
\end{equation*}
where $\lambda_{C}$ is the Compton wavelength of the electron that $\lambda
_{C}=\frac{\hbar}{m_{e}c}$. In this case, the eEDM increases with $\sqrt
{B_{0}}$. Oppositely, in the case of $\varepsilon\ll\sqrt{2eB_{0}\hbar c}$,
the eEDM expression becomes
\begin{equation*}
p^{1}\simeq e\lambda_{C}\frac{\varepsilon}{\sqrt{2eB_{0}\hbar c}},
\end{equation*}
and the eEDM decreases with $\frac{1}{\sqrt{B_{0}}}$. And from the condition of $\frac{d^{2}p^{1}}{dB_{0}^{2}}<0$, we see that the eEDM takes maximum value in the $B_{0}=\frac{(2n+k+1)\varepsilon^2}{2ec\hbar(k+1)^2}$. To gain insight into the electrical dipole moment  expression, we plot it according to the magnetic field, B, and the kinetic energy, $\varepsilon$, in the following Fig.(1-2):
\begin{figure}[htp]
\begin{center}
\includegraphics[width=0.8\textwidth, angle=0]{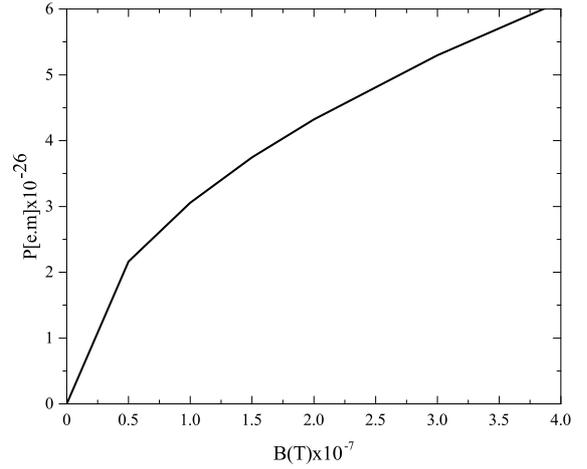}
\end{center}
\caption{Calculated eEDMs under constant magnetic fields by using
Eq.(\ref{Equation18}). In this calculation, we assume that an electron being in a $^{133}$Cs atom has $\varepsilon=2.6\times10^{5}eV$  and it undergoes transition from $6S_{1/2}$
to $6P_{3/2}$.}
\label{fig1}
\end{figure}
\begin{figure}[h]
\begin{center}
\includegraphics[width=0.8\textwidth, angle=0]{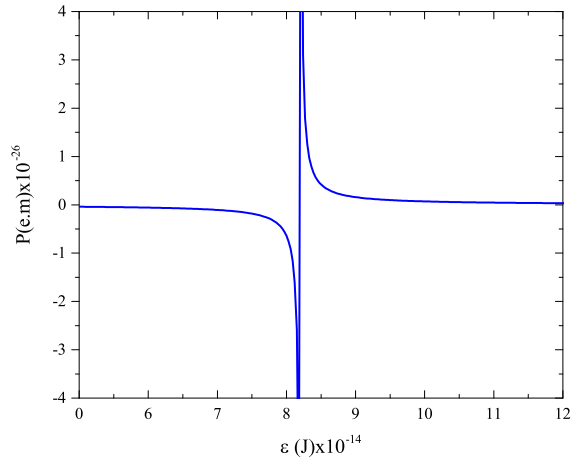}
\end{center}
\caption{Calculated eEDMs under constant magnetic fields by using Eq.(\ref{Equation18}). In this calculation,  by using $E-m_{e}c^{2}$, the kinetic energy and magnetic field values for the same  transition and atom , respectively , $\varepsilon=[10^{-14}, 10^{-12.5}J]$ and $ B= 10^{-10}T$ . }
\label{fig2}
\end{figure}

In order to indicate the completeness of our approach and behaviour of eEDM,
we first calculate the dipole moment by using the Eq.(\ref{Equation18}) under constant magnetic field for an
electron being in a $^{133}$Cs atom. For this calculation $6S_{1/2}%
\longrightarrow6P_{3/2}$ transition and experimental values in Ref.[24] are
used for kinetic energy of an electron and magnetic field. Here, we assume that
electron has a kinetic energy of $\varepsilon=2.6\times10^{5}eV$ and external
magnetic field becomes in the range of $B_{0}=[0,4]\times10^{-7}T$. Fig.1 shows
that as the magnetic field strength increases, the eEDM slightly increases and
does not change dramatically and it is also in the order of magnitude of
$10^{-24}$ with unit of (e.cm). It is obviously seen that our calculation for
above example is very close to previous findings[24].
\begin{table}[h]
\caption{eEDM values for different  magnetic fields and the kinetic
energy values of an $e^{-}$ in $^{133}$Cs atom.}%
\label{T1}
\begin{center}
\resizebox{7cm}{!} {
\begin{tabular}{l|c|c}
\hline\hline
\rule{0pt}{2.5ex}$\varepsilon= (E-m_{e}c^{2})$ & \quad${B_{0}}$ & $eEDM$ \\
\qquad$(eV)$ & \quad$(G) $ & \quad$(e.cm)$ \\
\hline\hline
\rule{0pt}{2.5ex}a)$\,10^{10}$ & $ [1,6]\times 10^{-3} $ & $3\times10^{-28} $ \\
b)$\, 10^{10}$ & $[1,6]\times 10^{2} $ & $9\times10^{-26} $ \\
c)$\, 10^{5} $ & $[1,6]\times 10^{3} $ & $3\times 10^{-20}$ \\
d)$\, 10^{9} $ & $[1,6]\times 10^{3} $ & $3\times10^{-24} $ \\
\hline\hline
\end{tabular}}
\end{center}
\end{table}

We can also extend our calculations to different values of kinetic energy of
an electron being in a $^{133}$Cs atom and external magnetic fields for
$6S_{1/2}\longrightarrow6P_{3/2}$ transition. We first keep the $\varepsilon$
constant at $10^{10}$ eV while gradually increasing the $B_{0}$. Then, the $B_{0}$
is kept constant at $6.0\times10^{-1}$ while increasing the $\varepsilon$.
Using the values for $\varepsilon$ and $B_{0}$ given in Table I, we obtain
the large number of values for eEDM as shown in third column of Table I.
\begin{table}[h]
\caption{Comparison of  our approach with previously published experimental studies. }
\label{T2}
\begin{center}
\resizebox{9.4cm}{!} {
\begin{tabular}
[c]{c|l|c||lc}\hline\hline
\rule{0pt}{2.5ex} $\varepsilon= (E-m_{e}c^{2})$ & \qquad${B_{0}}$ & $eEDM$ $in$
$this$ $work$ & \qquad$Bound$ $for$ $\left\vert d_{e}\right\vert $ & \\
\qquad$(eV)$ & \qquad$(G)$ & \quad\quad$(e.cm)$ & \qquad\qquad\quad$(e.cm)$ &
\\\hline\hline
\rule{0pt}{3ex}$2.6\times10^{5}$ & $1.0\times10^{-3}$ & $3.1\times10^{-24}$
& $\left\vert d_{e}\right\vert \leq7.7\times10^{-22}$ & [24]\\
$6.0\times10^{6}$ & $2.0\times10^{-6}$ & $5.9\times10^{-27}$ & $\left\vert
d_{e}\right\vert $=$1.3\times10^{-29}$ & [25]\\
$1.0\times10^{7}$ & $2.55\times10^{-2}$ & $2.3\times10^{-25}$ & $\left\vert
d_{e}\right\vert $=$(2.7\pm8.3)\times10^{-27}$ & [26]\\
$1.0\times10^{7}$ & $7.0\times10^{-3}$ & $2.1\times10^{-25}$ & $\left\vert
d_{e}\right\vert $=$4.0\times10^{-27}$ & [27]\\
$0.1\times10^{7}$ & $9.0\times10^{-3}$ & $2.0\times10^{-25}$ & $\left\vert
d_{e}\right\vert \leq1.6\times10^{-27}$ & [28]\\\hline\hline
\end{tabular}}
\end{center}
\end{table}

In order to check the validity of our approach, we compare our calculations
with previously published researches. In our all calculations, reported values for $B_{0}$, $\varepsilon$ and
related transition for an atom are considered. For example, eEDM is, first, calculated
for $B_{0}$ $=$ $2.0\times10^{-10} T$ and $\varepsilon$ $=$ $6.0\times10^{6}eV$,
and it is obtained as $5.9\times10^{-27}e.cm$ for $6^{2}S_{1/2}\longrightarrow
7^{2}P_{1/2}$ transition in $^{205}Tl$ atom. Secondly, we use
$0.255\times10^{-4}T$ and $1.0\times10^{7}eV$ values for $B_{0}$ and
$\varepsilon$, respectively. For $6^{2}P_{1/2}\longrightarrow7^{2}S_{1/2}$
transition in $^{205}Tl$ atom, the eEDM's magnitude is calculated as
$2.3\times10^{-25}e.cm$. Moreover, for $7.0\times10^{-7} T$ of $B_{0}$ and
$1.0\times10^{7}eV$ of $\varepsilon$ values, the eEDM's magnitude is found as
$2.1\times10^{-25}e.cm$ for $6P_{1/2}\longrightarrow6P_{3/2}$ transition in
$^{133}Cs$ atom. In our last calculation for $9.0\times10^{-7}T$ of
$B_{0}$ and $0.1\times10^{7}eV$ of $\varepsilon$ for $^{205}Tl$ in its
$6^{2}P_{1/2}$ ground state, the eEDM's magnitude is obtained as
$2.0\times10^{-25}e.cm$. Our above calculations and previous findings are
summarized in Table II. It can be clearly seen from the Table II that for each
calculation we find eEDM values being very close to formerly reported values
in the range of ignorable error.

\section{Conclusion}

In this study, we present  an approach to compute the electrical and magnetic dipole moments of the electron by integrating the polarization and magnetization densities of the electron current on the hyper surface, respectively. The dipole moment expressions obtained by the approach are in compatible with the current experimental results. We also see that the magnetic field  has an important effect on the electrical dipole moment of the electron: At first the eEDM increases with the square root of the magnetic field, $\sqrt{B_{0}}$, until it reaches to a maximum value for a certain value of the magnetic field, and then it decreases with $\frac{1}{\sqrt{B_{0}}}$. It is important to note that no matter how less its value gets it never reaches to the value (i.e. $10^{-38}$e.cm) predicted by the standard model, even if the magnitude of the magnetic field is in the Planck scale (i.e. the eEDM $\rightarrow 10^{-34}$e.cm as $B_0\rightarrow 10^{53}$T). From these results in the study, we clearly see that, in the case of $\varepsilon\ll\sqrt{2eB_{0}\hbar c}$, the magnetic field, $B_0$, squeezes or focuses the charge distribution of the electron.

\section*{Acknowledgement}
The Authors thanks Nuri Unal, Timur Sahin and Ramazan Sahin for usefull discussion.

\bibliographystyle{amsplain}

\end{document}